\documentclass{aa}
\usepackage{graphicx}
\usepackage{amssymb}
\usepackage{txfonts}
\begin{document}

\title{The progenitor of the 'born-again' core V605~Aql and the relation to its younger twin V4334~Sgr\thanks{Based on observations made at ESO, LaSilla Chile}}
\titlerunning{The progenitors of V605~Aql and V4334~Sgr}

\author{M.F.M. Lechner \and S. Kimeswenger}

\offprints{M.F.M. Lechner,
\\ \email{michaela.lechner@uibk.ac.at}}

\institute{Institut f{\"u}r Astrophysik der Universit{\"a}t
Innsbruck, Technikerstr. 25, A-6020 Innsbruck, Austria }

\date{Received 5. March 2004 / Accepted  22. June 2004}

\abstract{ We derived the properties of V605~Aql before the final
helium flash pulse by studying its surrounding PN A58.
Photoionizing models of our spectral data together with a new
distance estimate and a closer look at the recombination
timescales lead to a consistent model. Comparing our findings with
the only hydrogen-poor twin, namely Sakurai's Object, we conclude
that these born-again objects have normal PNe core masses. We are
able to prove V605~Aql indeed to be, similar to V4332 Sgr, a very
late thermal pulse object and to put constraints for the
evolutionary time scales for the transition back to the AGB.

  }

\maketitle


\section{Introduction}

Calculations done in the early 1980s demonstrated that luminous
central stars of PNe could be undergoing a {\it final helium shell
flash} before the final cooling of the white dwarf. It was
reviewed by (Iben \& MacDonald \cite{ibenmac}) that this {\it last
thermal pulse} brightens a white dwarf to AGB luminosity and
results in a sudden change in abundances to heavier elements as
the residual hydrogen envelope is consumed in the helium-burning
convective shell. Back at the AGB the central star of the PN
(CSPN) pursues again its evolutionary path to become a white
dwarf, but this time the main energy source is helium burning.
This evolutionary behavior is called the {\it born-again
scenario}. Born-again PNe helium-flash objects are characterized
by highly processed hydrogen-deficient nebular material and dust
and are of fundamental importance for the understanding of the
late stages of stellar evolution. Due to the short timescales of
the transition only few born again objects were discovered
therein. At the moment there are actually only three objects -
Sakurai's Object (V4334~Sgr), V605~Aql and FG~Sge -, where we can
observe more than a final H-deficient PN. This makes individual
detailed studies of these objects necessary, especially as they
are also discussed as possible progenitors of [WC]-CSPNe, PG1195
type CSPNe and RCrB stars. For a review see Clayton
(\cite{clayton01}) and references therein.

Sakurai's Object (V$4334$~Sgr) is the proto-typical final helium
shell flash object and had its flash in 1996 (D\"urbeck \& Benetti
\cite{duerbeck}). The old PN of V$4334$~Sgr has been found and its
spectrum reveals it to be an ordinary PN (Pollacco
\cite{pollacco99}).

V$605$~Aql is often referred to as the 'older twin of Sakurai's
object', as many similarities exist. It was discovered in 1919 as
'Nova Aquilae No.4' and later assigned V$605$~Aql. Only in the
1980s was it recognized to be a final helium flash object and thus
it was observed less than Sakurai's object or FG Sge. It had its
flash already in 1917, and therefore serves as pathfinder for
future events in V$4334$~Sgr, which is still frozen in its reborn
AGB resp.~post-AGB state. A detailed historical review of the
evolution of V$605$~Aql can be found in Clayton \& de Marco
(\cite{claytondemarco}) and the photometric similarities between
V$605$~Aql and V$4334$~Sgr are discussed in D\"urbeck et al.
(\cite{duerbeck02}).

Different theoretical models exist that should describe the
born-again phenomenon, but there is not yet a consent about the
crucial parameters like convective mixing, opacities at high
ionization energies, the importance of convective overshoot or
surface cooling by means of stellar wind (e.g. Lawlor \& MacDonald
\cite{lawlor}, Herwig \cite{herwig}).

We present here a detailed study of V605~Aql and compare our
results to those found for V4334~Sgr.

\section{Observations and Data Reduction}

We obtained spectra using the ESO NTT telescope at LaSilla (August
4$^{\rm th}$ to 6$^{\rm th}$ 2002) with the multi mode instrument
EMMI mounted. Also different narrow-band filter images were
obtained there. The calibration (bias, flatfield, wavelength
calibration and response curve for the spectra; bias and skyflats
for the images) was done using usual procedures in MIDAS. The
resolutions of the individual grisms used for the spectra can be
found in Table~\ref{grisms}.

\begin{table}[htbp]
\caption{Grisms used for spectroscopy} \label{grisms}
\begin{tabular}{lcr}
\hline
Grism& Resolution &\hbox{\ $\lambda$\ \ } \\
\hline \hline
grism \#2& ~~ 0.35~nm/pixel ~ & ~ 390 to 980 nm \\
grism \#3& ~~ 0.29~nm/pixel ~ & ~ 390 to 850 nm \\
grism \#6& ~~ 0.15~nm/pixel ~ & ~ 500 to 880 nm \\

\hline \hline
\end{tabular}
\end{table}

\begin{figure} [ht]
\centerline{\resizebox{8cm}{!}{\includegraphics{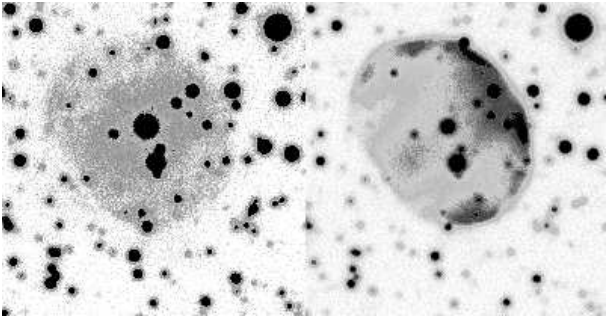}}}
\caption{These [OIII] (left) and [NII] (right) narrow-band images
of PN A58 are not rotated to prevent information loss. The
position angle to a north-east orientation is $132.381$. The edge
enhancement in [NII] gives the impression of an interaction with
the ISM.}\label{narrowbands}
\end{figure}

Narrow-band direct images (see Fig. \ref{narrowbands}) show a
strong dependence of the shape of PN A58 on the excitation level.
The [OIII] image of the PN looks rather roundish, whereas in
H$\beta$ and in [NII] a western rim is clearly visible. In all
three images the nebula seems to consist of three parts; an inner
central part and two 'polar caps', which is most obvious in the
[OIII] image. According to the manual the EMMI CCD chip has a
resolution of 0.\arcsec333 per pixel, but the seeing was not
sufficient to derive the size of the V$605$~Aql knot as done by
Hinkle et al. (\cite{hinkle01}). On the basis of the CCD images
our estimate for the size of the old PN A58 is 50\arcsec \, for
the longitudinal axis and 38\arcsec\, for the secondary axis.

The edge enhancement suggests an interaction with the interstellar
medium (ISM). Hence we determined the physical properties of the
old nebula Abell~58 out of the inner part, nearer to V$605$~Aql.

The interstellar extinction had been derived from the Balmer
decrement. Different regions show an $E(B-V)$ with extrema of
$0\fm35$ and $0\fm55$. A weighted average of $E(B-V)=0\fm5$ was
adopted for the models.

\section{Distance Determination}
\label{distance}

There were several distance estimates for V$605$~Aql before,
giving values between 2.7 and 6.0 kpc as reviewed by Clayton \& de
Marco (\cite{claytondemarco}), who recommended a distance of 3.5
kpc.

We tried to determine the distance by the use of radio
observations. The VLA Sky Survey by (Condon \& Kaplan
\cite{condon}) gives a $S=2.7$~mJy flux-density peak for PN A58 at
1.4~GHz (according to a wavelength of 21~cm), with an 0.8~mJy RMS
uncertainty. Usually distances are calculated by the 6~cm radio
fluxes (according to a frequency of 4885~MHz), but assuming
free-free radiation with
$$I=I_0 \,\nu^{-0.1}$$
we translated this value into
$$S_{6cm}=0.0024\pm0.0007~\rm Jy\, .
$$
Now the method by Cahn et al. (\cite{cks}) can be applied. The
optical thickness parameter $\tau$ is given by
$$\tau=\log\left(\frac{4\theta^2}{S_{6cm}}\right) \,,$$
with $S_{6cm}$ in Jansky and $\theta$ in arcseconds. A mean value
for the angular radius of the PN of 22\arcsec\ resulted in $\tau =
5.91\pm0.12$. For $\tau>3.13$ the ionized mass $\mu$ of the PN can
be described with
$$\log(\mu)=-0.87 \,.$$
Using these equations, the radius $R$ and the distance $D$ of the
PN can be determined from the ionized mass by
$$\log(R)=\frac{\log(\mu)}{2.5}-1.306+\frac{\tau}{5}=-1.654+\frac{\tau}{5}$$
and
$$D=206.265 \, \frac{R}{\theta} \,.$$
So the solutions using (Condon \& Kaplan \cite{condon}) and (Cahn
et al. \cite{cks}) are
$$R=0.34\pm0.02 \rm~pc$$ and $$D=3.15\pm^{0.23}_{0.15} \rm~kpc\,
.$$

Alternatively the method of van de Steene \& Zijlstra
(\cite{steene}) uses the radio continuum brightness temperature
$$T_b=\frac{c^2}{2 \pi k \nu^2} \frac{S_{6cm}}{\theta^2}=18400 \,  \frac{S_{6cm}}{\theta^2} = 9.05\times10^{-2} \,.$$
From this temperature, which is a measure for the brightness of
the nebula at radio frequencies, together with their fitted linear
regression line through a sample of bulge PN they specify a radius
$R$:
$$\log(R)=-0.35(\pm0.02)\, \log(T_b) - 0.51 (\pm0.05)$$
leading to $R=0.71$~pc and $D=6.6$~kpc. This approximation is
generally a factor of 2 too big.

Normally the best and most faithful results (for a discussion see
Schmeja \& Kimeswenger \cite{schemimex}) are derived by the method
of Schneider \& Buckley (\cite{schneiderbuckley}). Their equation
is:
$$\log(D)=-\log(\theta) -0.0261\log^2(I) - 0.299\log(I) + 1.198 \,,$$
where the surface brightness $I=S_{6cm}/(\theta^2 \pi)$ is
measured in mJy arcsec$^{-2}$. This leads to
$D=3.08\pm^{0.17}_{0.12}$~kpc.

For all further calculations we therefore assumed a distance of
$D=3.1$~kpc. This means that the radius $R$ of the PN A58 equals
$0.33$~pc.

\section{Photoionization Models}
\label{cloudy}

On the basis of the line fluxes presented in Table \ref{pnspec}
the old PN A58 was modelled with CLOUDY (Ferland \cite{ferland}).
The final goal was not to reproduce the ionization and thermal
equilibria within the nebula in detail but to qualify the stellar
continuum before the helium flash pulse. The low density PN sees
still its old central star at the time, when it was sufficiently
hot to ionize it. Thus a large grid of photoionizing models was
calculated, giving us the position of V$605$~Aql in the
Hertzsprung-Russell diagram (HRD) during its hidden past.

\begin{table}[htbp]
\caption{Dereddened line fluxes for the PN A58} \label{pnspec}
\begin{tabular}{lrrl}
\hline
Line ID & \hbox{\ $\lambda$\ [\AA] } & Intensity [H$\beta$ = 100]&\\
\hline \hline
HeII &   4686    & 27 &\\
H$\beta$ & 4861 & 100 &\\
OIII  & 4959 & 286&\\
OIII  & 5007 & 804&\\
NII &  5755    & 4.5 & $\pm1.0$\\
HeI &  5876    & 18 &\\
NII &  6548    & 92 &\\
H$\alpha$ & 6563 & 286 &\\
NII &  6584    & 280 & $\pm20.0$\\
SII &  6716    & 42 &\\
SII &  6731    & 31 &\\
ArIII &  7136    & 13 &\\

\hline \hline
\end{tabular}
\end{table}

Usually blackbody radiation is used for such calculations of other
objects (e.g. van de Steene \cite{steenethesis}, van Hoof \& van
de Steene \cite{hoofsteene}, Kerber et al. \cite{kerber99}, N{\"
u}rnberger et al. \cite{nurnberger}, Cuisinier et al.
\cite{cuisinier}). Armsdorfer et al. (\cite{birgitmexiko}) and
Armsdorfer \& Kimeswenger (\cite{birgit}) showed that the use of
real stellar atmospheres changes especially the helium lines. Here
we thus used NLTE central star models of Rauch (\cite{rauch97},
\cite{rauch2003}). The same grid of model atmospheres was used
also by Pollacco (\cite{pollacco99}) for the twin object
V4334~Sgr.

The temperature of the central star of the PN, its luminosity, the
gas density and the filling factor of the gas were varied. While
the range for the filling factor was selected to fit to the
typical findings for evolved nebulae (Armsdorfer et al.
\cite{birgitmexiko}), the gas density range of 100 to 300
e$^-$cm$^{-3}$ was taken from the ratio of the [SII] doublet as in
Osterbrock (\cite{osterbrock}). The distance of 3.1$\,$~kpc was
adopted from section \ref{distance}, leading with the average
angular radius of 22\arcsec \, to the linear radius of the shell.

\begin{figure}[ht]
\centerline{\resizebox{8cm}{!}{\includegraphics{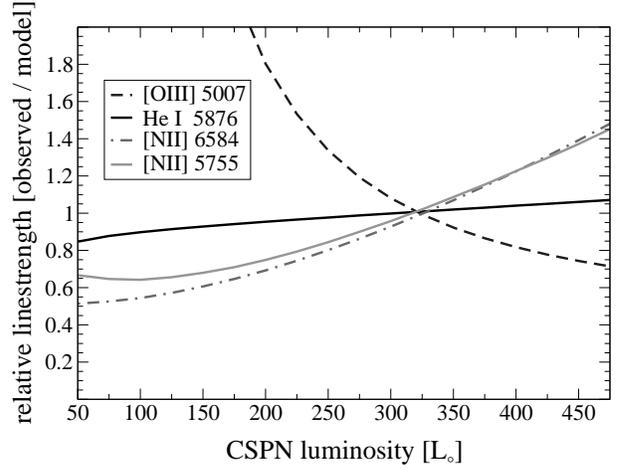}}}
\caption[Best Fit Model]{The best fit model shows a good
intersection of the solutions for the Helium, Oxygen and Nitrogen
lines at the level $I_{\tt observed}/I_{\tt model} = 1.0$. The
resulting model is given in Table \ref{resulttab}.} \vspace{-3mm}
\label{bestfit.eps}
\end{figure}

The strength of the helium lines is nearly independent from the
CSPN luminosity. Thus this fixes the abundance of this element
clearly. The variations of the parameters were chosen to get a
optimized intersection of the solutions for the HeI$_{5876}$, the
[OIII]$_{4948+5007}$ and the [NII]$_{6548+6583}$ lines at the
level of $I_{\tt observed}/I_{\tt model} = 1.0$. These solutions
are not completely independent from the chosen abundance. The
[NII]$_{5755}$ line then should intersect, if the resulting
electron temperature is suited, there as well. This fixes the CSPN
temperature and therefore the whole solution. The best fit model
is shown in Fig. \ref{bestfit.eps}.

\begin{table}[htb]
\caption{Parameters of the best fit model.} \label{resulttab}
\begin{tabular}{l l}
\hline
Parameter  &  Value\\
\hline \hline
distance & 3.1\,kpc (see section \ref{distance}) \\
{\it CSPN}: & \\
~~~~luminosity & $L = 325\pm$100\,L$_\odot$ \\
~~~~effective temperature & $T_{\rm eff} = 121.5\pm6\,{\rm kK}$ \\
~~~~age from post-AGB & $t = 8300^{+1500}_{-500}$\,\,years \\
~~~~white dwarf mass & $M = 0.605\pm0.02$\,\,M$_\odot$ \\
{\it PN gas shell}: & \\
~~~~average radius & angular 22\arcsec \\
~~~~& linear $r = 1.02\,\,10^{16}\,{\rm m}$ \\
~~~~thickness & $\Delta_r/r = 0.2$ \\
~~~~filling factor & $\epsilon = 0.05$ (adopted)  \\
~~~~electron density & $n_e = 200\pm100\,\,{\rm e}^-{\rm cm}^{-3}$ \\
~~~~abundances: & \\
~~~~~~helium & [He] = $-$0.83 \\
~~~~~~nitrogen & [N] = $-$4.00 \\
~~~~~~oxygen & [O] = $-$3.40 \\
\hline \hline
\end{tabular}
\end{table}

\begin{figure}[ht]
\centerline{\resizebox{8cm}{!}{\includegraphics{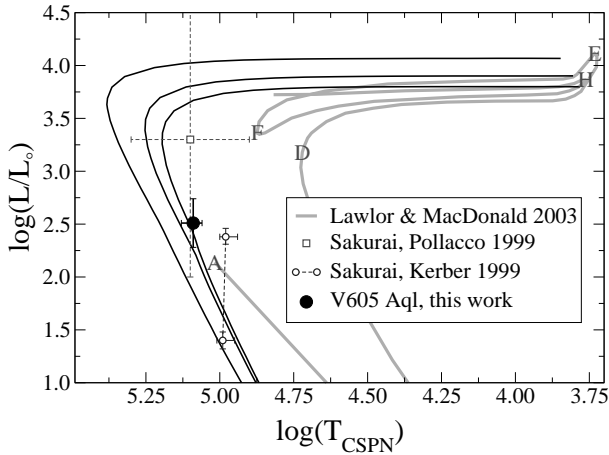}}}
\caption[Position of the CSPN in the HRD]{The position of the CSPN
in the HRD at the moment of the helium flash. The evolutionary
tracks for the post-AGB evolution for a 0.605, 0.625 and
0.696\,M$_\odot$ white dwarf from Bl\"ocker (\cite{bloecker}) are
plotted. For comparison reasons also the results of Pollacco
(\cite{pollacco99}) and of Kerber(\cite{kerber99}) for Sakurai's
Object are shown. Pollacco's error bars show the observational
incertainties, he declares $log(L/L_{\sun})<3$ to be consistent
with the white dwarf cooling tracks. Kerber assumed a relative
error of 10 percent in the intensities, but admits that the
overall error bars are certainly larger. The grey line underneath
represents the evolutionary model of Lawlor \& MacDonald
(\cite{lawlor}).} \vspace{-3mm} \label{hr.eps}
\end{figure}

The placement of the derived CSPN luminosity and temperature
together with the theory of stellar post-AGB evolution by
Bl\"ocker (\cite{bloecker}) yields the mass of the final white
dwarf and the transition time $t$. The position of the CSPN in the
HRD is shown in Fig. \ref{hr.eps}. The resulting model (see Table
\ref{resulttab}) is completely self consistent. In addition it is
also supported by the velocity measurements by Pollacco  et al.
(\cite{pollacco92}), who derived for the expansion velocity of A58
$v_{exp}=31\pm4$~km\,s$^{-1}$, giving with the dynamical age of
8200 years a dynamical radius of PN A58 of
$$R=0.26\pm^{0.04}_{0.05} {\rm pc} \, ,$$
which matches extremely well our radius of $R=0.33~{\rm pc}$ as
derived in section \ref{distance}. This is an especially good
conjunction, if we take into account that the dynamical solution
is usually a factor of 2 or 3 off target.

\section{Recombination Timescales in A58}
\label{recomb}

There is definitely HeII$_{4685}$ in the spectrum of A58. This
fact provides together with the knowledge of recombination
timescales an estimate for the time between the ending of the UV
radiation of the central star and the nova burst (position {\tt H}
or {\tt E} in Fig. \ref{hr.eps}).

Using the same dielectric recombination coefficients $\alpha$ as
Pollacco (\cite{pollacco99}) we calculated a table for the
recombination time-scales for electron densities $n_e$ of 100 and
$200$~e$^-$cm$^{-3}$ (see Table \ref{recombitab}).

\begin{table}[htbp]
\caption{Recombination time-scale for atomic states at
$T_e=10.000$~K for electron densities $n_e$ of 100 and
$200$~e$^-$cm$^{-3}$. An increase of the electron density to
$n_e=300$~e$^-$cm$^{-3}$ only lowers the upper limits. The terms
'r' and 'd' refer to the radiative and dielectronic components in
the total recombination coefficient 't'.} \label{recombitab}
\begin{tabular}{lrr}
\hline
State & $\tau_r(n_e=100)~[yr]$ & $\tau_r(n_e=200)~[yr]$\\
\hline \hline
H$^0$ &   760~~~    & 380~~~\\
He$^0$ & 1500~~~ & 750~~~\\
He$^+$  & 206~~~ & 103~~~\\
O$^{2+}$(r)  & 31~~~ & 16~~~\\
O$^{2+}$(d)  & 28~~~ & 14~~~\\
O$^{2+}$(t)  & 19~~~ & 9~~~\\
O$^+$(r)  & 155~~~ & 78~~~\\
O$^+$(d)  & 190~~~ & 95~~~\\
O$^+$(t)  & 85~~~ & 43~~~\\
N$^+$(r)  & 139~~~ & 70~~~\\
N$^+$(d)  & 155~~~ & 78~~~\\
N$^+$(t)  & 73~~~ & 37~~~\\
\hline \hline
\end{tabular}
\end{table}

It can be seen that the recombination times for [OIII] are
relatively low, but this table is only valid for single atomic
systems. We have to take into account the Bowen HeII Lyman
$\alpha$ - [OIII] resonance-fluorescence mechanism within the PN.
This mechanism slightly decreases the HeII recombination times but
enlarges the [OIII] recombination time-scales. For a more detailed
study refer to Osterbrock (\cite{osterbrock}). The emphasis lies
upon HeII as most common element, but also NII is important as it
is not influenced by the Bowen HeII  - [OIII]
resonance-fluorescence mechanism.

Since we have to take the minimum of all the derived time-scales
for a specific electron density as an upper limit for the lifetime
of a nebula in this state, we can use a weighted average of
$100\pm50$ years as a good estimate for the upper limit. This also
justifies the use of photoionizing models with CLOUDY.

\section{Discussion and Conclusion}

The parameters used to calculate stellar evolution become crucial
when we consider such subtleties as born-again objects. Hence the
physical properties of born-agains in all stages provide a hard
constraint on valid evolution theories. In general two different
scenarios are discussed for the {\it
born again} event:\\
\begin{itemize}
\item The late thermal pulse (LTP) where the helium flash occurs
while the star is still H-shell burning. The CSPN has still a
luminosity above a few thousands L$_\odot$. H-ingestion to the
flash convection zone causes -- according to current models -- a
slow evolution back to the AGB lasting at least several hundreds
of years (Bl\"ocker \& Sch\"onberner \cite{bloecker_fgsge}).

\item The very late thermal pulse (VLTP) where the CSPN was
already on the white dwarf cooling track, i.e., after the
cessation of H burning. In this scenario the stellar luminosity
decreases slowly from the flash position (position {\tt A} in
Fig.~\ref{hr.eps}) to a very low luminosity (called {\tt B} in
Lawlor \& MacDonald \cite{lawlor}). The timescale for this
transition is discussed very controversially. While Lawlor \&
MacDonald have transition timescales of up to several thousands of
years, Iben et al. (\cite{iben83}) and Iben \& MacDonald
(\cite{ibenmac}) give a few tens of years, Herwig et al.
(\cite{herwig99}) and Herwig (\cite{herwig}, \cite{herwigIAU})
give a timescale of less than one year for a 0.604~M$_\odot$ CSPN.
Older models give such a fast evolution only for high mass CSPNe.
\end{itemize}
For a more detailed discussion see Bl\"ocker (\cite{ltp_vltp}) and
references therein.

 Based on recombination timescales we conclude that the time
between the end of the UV radiation of the central star
($t_{flash} =$ the time of the late helium flash) and the observed
bright Nova burst of V$605$~Aql in the 1920's ($= t_{nova}$) has
to be very short. Modelling the surrounding old PN A58 with
CLOUDY, we obtain that the central star had a mass near
0.6~M$_\odot$ (using the tracks of Bl\"ocker (\cite{bloecker}) as
well as those of Herwig et al. (\cite{herwig99}) and Herwig
(\cite{herwig}) ) and a temperature of 120\,000 Kelvin before the
final helium flash. The model of Herwig (\cite{herwig}) with lower
masses is neither consistent with the photometric data of V4332
Sgr, as pointed out there already, nor with our model of the
pre-outburst CSPN of V605 Aql. According to Bl\"ocker
(\cite{bloecker}) the time since leaving the post-AGB resulted in
8200 years. The distance of $3.1\pm{0.2}$~kpc from radio
observations goes well with the constraints by the photoionization
model and also matches the angular size for the derived age. Our
observations obtained in 2002 ($= t_{obs}$) and thus about 85
years after $t_{nova}$ and the finding that the age of the PN
($t_{obs}-t_{flash}$) has to be around 100 years, lead to the
conclusion that the transition time to either point {\tt H} or
{\tt E} has to be below 20 years - taking into account all types
of uncertainties. This clearly excludes a LTP and accordingly is
consistent with the CSPN position in the HRD. Our findings clearly
- both in luminosity and especially in temperature - exclude the
possibility that the PN was excited in position {\tt D} or {\tt F}
(see Fig.~\ref{hr.eps}). In addition, test calculations with
CLOUDY also exclude the possibility that the PN was excited
additionally in between by a $40\,000\,<\,\mbox{\rm T}_{\mbox{\rm
CSPN}}\,<\,70\,000\,$K source with a luminosity more than
1000\,L$_\odot$. This would have imprinted by special emission
line ratios of [NII] and [SII]. Also, although checked by Fuhrmann
(\cite{fuhrmann}) on more than 430 Sonneberg and Heidelberg patrol
plates taken after 1920 and by D\"urbeck et al.
(\cite{duerbeck02}) on the Harvard patrol plates, no brightening
has been detected since the 1917 outburst. This implies that the
source is either in its first return to the AGB (position {\tt
E}), or not visible due to a thick dust shell during the
transition {\tt E-F-H}, or this second loop does not really occur.
Whilst Iben et al. (\cite{iben83}) do not show such a loop at all,
the transition time for the model of Lawlor \& MacDonald
(\cite{lawlor}) gives timescales of only a few tens of years for
the heating process after the first "Nova event" for the
transition {\tt E-F}. The timescale in Herwig (\cite{herwigIAU})
is even shorter. Then it stays near {\tt F} for a long time at
high luminosity and high temperature. This environment would have
destroyed a dense dust shell built during the short evolution from
leaving {\tt E} towards {\tt F} (Koller \cite{joemsc}, Koller \&
Kimeswenger \cite{joe}, \cite{joeapj}, Kimeswenger \& Koller
\cite{nilfisc_sakurai}) and consequently V605 Aql should have been
observable as a blue star on the POSS plates. This wasn't found
neither on the first generation POSS plates in the 1950s nor on
the second generation in the 1980s and 1990s. We thus conclude
that the object is at its first return to the AGB and the
timescale for the transition has to be as short as given by Herwig
(\cite{herwig}) and not as long as given by Lawlor \& MacDonald
(\cite{lawlor}). Furthermore V605~Aql should have stayed at least
several decades at the AGB before reheating to the temperature
seen nowadays in the CIV lines (Kimeswenger \cite{RevMex}, Lechner
\& Kimeswenger \cite{ANS}). This gives somewhat longer timescales
than those predicted by Lawlor \& MacDonald (\cite{lawlor}) or by
Herwig (\cite{herwigIAU}).

The mass of the progenitor CSPN also was discussed
controversially. D\"urbeck et al. (\cite{duerbeck97}) derive
$\geq$0.7~M$_\odot$ using Paczynski's core-mass relationship.
D\"urbeck et al. (\cite{duerbeck2000}) argue, using the high
luminosity and the evolution time span, that a mass of
0.8~M$_\odot$ should be a lower limit. However the outcomes here
for V605~Aql and those of Pollacco (\cite{pollacco99}) and Kerber
(\cite{kerber99}) (see Fig. \ref{hr.eps}) place V$605$~Aql and
V$4334$~Sgr clearly in the range of normal PNe core masses at or
just below 0.6~M$_\odot$. Kerber (\cite{kerber99}) calculated his
models with the photoionization code GWYN and used only blackbody
radiation. Pollacco (\cite{pollacco99}) used CLOUDY with different
but comparable stellar atmospheres. It can be assumed that their
spectra are from different regions in Sakurai's Object. In Fig.
\ref{hr.eps} we see also beautifully the dependence on the
distance with the aid of Kerber's results for 1.5 kpc and 5.5 kpc
for Sakurai's Object. The mass of 0.6~M$_\odot$ also corresponds
extremely well with the dynamic age of the old PN. A CSPN with a
mass above 0.7--0.8~M$_\odot$ would evolve much faster than the
observed PN. So we conclude that the CSPN mass is very similar to
that of Sakurai's object found by Herwig (\cite{herwig}).


\end{document}